\documentclass[12pt]{article}
\usepackage{geometry}
\usepackage{graphicx}
\usepackage{amsmath}
\usepackage{mathtools}
\usepackage[thinc]{esdiff}
\usepackage[bottom]{footmisc}
\usepackage{float}
\usepackage{graphicx}
\usepackage{dcolumn}
\usepackage{bm}
\usepackage{braket}
\usepackage{xfrac}
\usepackage{authblk}
\usepackage{blindtext}
\title{Low Energy Nuclear Reactions Through Weak Interactions }
\date{June 2024}
\author[1]{K. Ramkumar\thanks{kaanapuliramkumar08@gmail.com}}
\author[1]{Harishyam Kumar\thanks{hari@iitk.ac.in}}
\author[2]{Pankaj Jain\thanks{pkjain@iitk.ac.in}}
\affil[1]{{}Department of Physics, Indian Institute of Technology, Kanpur}
\affil[2]{{}Department of Space, Planetary \& Astronomical Sciences \& Engineering, Indian Institute of Technology, Kanpur}
\begin{document}
\maketitle

\begin{abstract}
	We consider the possibility that low energy nuclear reactions (LENR) arise due to the conversion of proton to neutron through 
weak interactions. 
The resulting neutron forms a short-lived virtual state, which then gets captured by another nucleus through photon emission. This whole process happens under the framework of second order perturbation theory. We find that the
	rate of this process is negligibly small in most cases. However, in the presence of a resonance the rate
	can be substantial and observable. 
\end{abstract}

\section{Introduction}
There is considerable experimental evidence that nuclear fusion reactions may take place at observable rate
even at very low incident energy \cite{PhysRevC.78.015803,doi:10.1002/9781118043493.ch41,doi:10.1002/9781118043493.ch42,doi:10.1002/9781118043493.ch43,biberian2020cold,StormsCS2015,McKubre16,Cellani19,Mizuno19,SRINIVASAN2020233,Huang09}.
Theoretically, despite several attempts, it has proved to be difficult to find a mechanism which can explain this phenomenon. 
The proposed mechanisms include, screening  contributions \cite{assenbaum1987effects,ichimaru1993nuclear,PhysRevC.101.044609},  
correlated states \cite{articleVy,PhysRevAccelBeams.22.054503}, electroweak interactions \cite{Widom10}, formation of 
nuclear clusters \cite{SPITALERI2016275}, electronic deep orbit solution to Dirac equation \cite{Meulenberg19} and 
phonon contributions \cite{Hagelstein15}.
Besides these, it has been suggested that at second order in perturbation theory, new 
fusion processes may open up which may not be as suppressed as the standard first order process
\cite{PhysRevC.99.054620,Jain2020,Jain2021,ramkumar2022,jain2024medium}. In this case the process takes place
by two interactions. The first interaction, which may take place due to the presence of another particle in medium
\cite{PhysRevC.99.054620,jain2024medium} or by photon emission \cite{Jain2020,Jain2021,ramkumar2022}, creates a 
virtual short-lived state. This is not an eigenstate of energy. It is formed by superposition of eigenstates of all
energies and hence energy conservation is not applicable at individual vertices, as is governed by the rules
of the second order perturbation theory. Of course, there is overall conservation of energy. 
Due to the contribution of eigenstates of all energies, the Coulomb barrier may not be prohibitive and hence it is possible that the fusion process can proceed at observable rates. However, it has been shown that even in this case the rates may 
be quite small unless some special conditions are met \cite{Jain2020,Jain2021,ramkumar2022,jain2024medium}.

In the present paper we consider a specific process which proceeds at second order in perturbation theory.
We consider fusion of a proton with a 
 nucleus X which has $Z$ protons and $A$ nucleons. The initial state
particles are assumed to have low energy of order eV. The Coulomb repulsion
between these two particles is prohibitive and the standard fusion process
would lead to extremely small reaction rates which are unobservable in
laboratory. 
We assume that the first interaction happens due to a weak process in which the incident proton is
converted to a virtual neutron with emission of a neutrino and a positron. 
Subsequently, the neutron
gets captured by the target nucleus $X$ through emission of a photon.
As mentioned earlier,
the intermediate nuclear state is not an eigenstate of the Hamiltonian 
and we need to sum over all eigenstates.
In all such second order processes, it has been observed that while the amplitude is large for some of these
eigenstates, the sum has a tendency to cancel and become very small \cite{Jain2020,Jain2021,ramkumar2022,jain2024medium}. 
The amplitude turns out to be large 
for eigenstates whose wave vector is approximately opposite to that of the particles emitted at the first 
vertex. Several mechanisms have been suggested to evade this cancellation. These include, presence of 
resonance in the nuclear spectrum \cite{Jain2020,Jain2021,ramkumar2022} and medium effects \cite{Jain2020,Jain2021,ramkumar2022,jain2024medium}. 
As we shall see in this paper, the weak interaction induced process also suffers from the same problem, i.e. the sum over
all states tends to cancel. We then examine the contribution of a resonance. We find that if a resonance is present
with suitable properties, the rate can be substantial. The mechanism appears to be such that it may be valid in general
and not just for the specific weak interaction process being considered in this paper.

\begin{figure}[H]
\begin{center}
\includegraphics[ clip,scale=0.8]{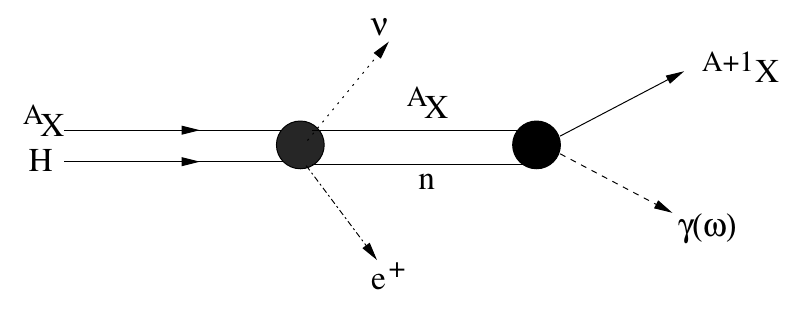}
\caption{\label{fig:process} Schematic illustration of the process given
	in Eq. \ref{eq:rxn}.}
\end{center}
\end{figure}

We consider the reaction of proton and a nucleus 
$X$ with atomic mass number $A$ to produce a nucleus with mass number $A+1$. 
As nuclear by products, positron, neutrino and high energy gamma photon are emitted. The process can be expressed as,
\begin{equation}\label{eq:rxn}
	{}^1{\rm H}\:+\:{}^AX\:\longrightarrow\: {}^{A+1}X\:+\:e^+\:+\:\nu\:+\:\gamma(w)
\end{equation}
For simplicity, we assume that initially the proton and ${}^AX$ form a 
molecular bound state. 
The two perturbations involved are: 
\begin{itemize}
	\item[1.] Conversion of proton to neutron through weak interactions
	\item[2.] Capture of neutron by ${}^AX$ with emission of a photon
 through electromagnetic interactions.
\end{itemize}
The process is schematically illustrated in Fig. \ref{fig:process}. 

\section{The H$-$X reaction at second order in  perturbation theory}\label{sec:amplitude}
We take the initial H$-$X molecular state to be s-wave ($l=0$). 
Here we are interested in displaying the basic 
mechanism which is likely to have a much
wider application and not confined to the particular process being
considered here. In particular, it may also be applicable to
cases in which the first interaction is not a weak process, as considered in Ref. \cite{Jain2020,Jain2021,ramkumar2022,jain2024medium}.
 As an example, we may consider X 
 to be the nickel nucleus with $A=58$. In this case,  
 the final state nucleus is nickel with $A=59$. As we shall argue, the same mechanism is applicable to a wide 
 range of nuclei that might have a sharp resonance at relatively low energy.
 We may treat the nuclear system
within the framework of the shell model, in which ground state of the nickel nucleus corresponds
to the $2p_{3/2}$ level with a binding energy of approximately 9 MeV. 
The intermediate state neutron in this case would have $l=0$.  

Let $\vec r_1$ and $\vec r_2$ be the position vectors of the ${}^A$X and  
proton respectively. Furthermore $\vec R =(m_1\vec r_1+m_2\vec r_2)/M $ and 
$\vec r =\vec r_2 - \vec r_1 $ denote the 
center of mass and relative coordinate respectively. Here $m_1$ and $m_2$
are the masses of particles ${}^A$X and the nucleon respectively and $M=m_1+m_2$. Here we shall treat proton and neutron as an isospin doublets, with proton as isospin up and neutron as isospin down. 
Hence the nucleon wave function is represented as
\begin{equation}
\Psi = \begin{bmatrix}
\psi_p \\ 
\psi_n
\end{bmatrix}
	\label{eq:isodoublet}
\end{equation}
in the isospin space. 
The parameter $m_2=m_N$ represents the mean mass of these two particles. We denote the neutron and proton mass difference as, $m_n-m_p =  \Delta m$. 
We split the Hamiltonian of the system, such that,
\begin{equation}
H = H_0 + H_I
\end{equation}
where $H_0$ is the unperturbed Hamiltonian and $H_I$ is the time 
dependent perturbation, which gets contributions both from weak and 
electromagnetic interactions. The unperturbed Hamiltonian is given by,
\begin{equation}\label{eq:h_0 weak}
    H_0\:=\:\frac{p_1^2}{2m_1} +{p_2^2\over 2m_2} - {1\over 2}\tau_3\:\Delta m\:c^2 +V
\end{equation}
where $V$ represents the potential energy term, 
including both the nuclear and molecular potential and $\tau$ are the
Pauli matrices, acting as generators in the isospin space. 
Here the kinetic energy part is taken to be isospin symmetric, with the
isospin breaking terms considered as part of the potential.  
The term in Eq. \ref{eq:h_0 weak} proportional to $\tau_3$ arises due to the mass difference between neutron and proton. This term can be justified by considering the non-relativistic limit of the Dirac equation, as shown in the Appendix. 

The initial state wavefunction can be expressed as
\begin{equation}
	\Psi_i(\vec r,\vec R) = \psi_i(\vec R) \chi_i(\vec r) |I_3=1/2>
\end{equation}
where $I_3$ refers to the third component of the isospin. Here we indicate
only the isospin of the nucleon state. 
We assume the center of mass wavefunction $\psi(\vec R)$ to be a plane wave.
This will not give any essential contribution and here we focus on the
relative coordinate. We assume that the system is in the $l=0$ ground state.
This state can be obtained by using an appropriate molecular potential. 
In application to LENR, we are interested in getting an order of
magnitude estimate to determine if the rate may be anywhere close to
being observable in laboratory. For this purpose, 
we simply use a Gaussian form of the wave
function peaked at the typical bond length for nickel hydrogen system, which
is expected to be about 3.7 atomic units. 
Hence we assume that $\chi_i(\vec r)$
takes the form
\begin{equation}
	\chi_i(\vec r) = {N_i\over \sqrt{4\pi}} \ e^{-(r - r_0)^2/\Delta^2}
 \label{eq:chir}
\end{equation}
Here $N$ is the normalization factor and we take $r_0= 3.5$ atomic
units, and $\Delta\approx 0.3$. The
value of $\Delta$ is dictated by the requirement that the wave function
is heavily suppressed as $r\rightarrow 0$ due to Coulomb repulsion. With
the choice of parameters, it is clear that this region would contribute
neglibly to the reaction rate. 

In order to compute the final state nuclear wave function we may assume the
shell model potential, 
\begin{equation}
        V_X(r) = {-V_0\over 1+\exp[(r-R_X)/a_X]}
        \label{eq:potentialX}
\end{equation}
where, $V_0=50 $ MeV, $a_X = 0.524$ fm and
$R_X = 1.25 A^{1/3}$ \cite{krane}. With these parameters
we obtain the energy eigenvalue for the 2p state of nickel to be $-9.2$ MeV. This
corresponds to the ground state of the ${}^{59}$Ni and is in good agreement
with the observed value of $-9$ MeV. 
We express the corresponding $l=1$ wave function as,
\begin{equation}
	\phi_{X,m}(\vec r) = {u_X(r)\over r  } Y_1^m
        \label{eq:wavefnX}
\end{equation}
Here we are interested in determining whether the rate may be anywhere 
close to being observable. For this purpose we may simplify our computation
by choosing a particular final state. Hence in our calculation we set $m=1$. 
Furthermore we take the spin state of the intermediate neutron and the initial
proton to be such that
$m_S=1/2$. 
This would correspond to a Fermi transition in which the nucleon spin does
not flip. 
The total spin quantum numbers of the final state nucleus 
${}^{59}$Ni in this case are $J=3/2$ and $m_J=3/2$.

The center of mass wave function of
the final state nucleus is taken to be a plane wave. The intermediate 
state wave function of the neutron-X system 
with $E>0$ and $l=0$ may be computed by using the
nuclear potential given in Eq. \ref{eq:potentialX}.  
It can be expressed as
\begin{equation}
\Psi_n(\vec r,\vec R) = \psi_n(\vec R) \chi_n(\vec r) |I_3=-1/2>
\end{equation}
As we shall see later, a non-negligible contribution is obtained only when the 
intermediate state is close to
a resonance. In order to study this, it is convenient to directly model the phase shifts
rather than use a potential model. The standard low energy expansion of the phase shift
$\delta$ for $l=0$, corresponding to the intermediate state, is given by
\begin{equation}
k\cot\delta = -a_1 + a_2 k^2 + ... 
\label{eq:reso}
\end{equation}
where we have not displayed higher order terms in $k^2$. Here $a_1$ and $a_2$ are parameters
related to the scattering length and effective range respectively. These can be obtained 
from the potential model. Using this expansion we can directly model the required 
wave function, as discussed in the next subsection.

The transition amplitude at second order in perturbation theory is given by,
\vspace{-7mm}
\begin{center}
\begin{multline} \label{eq:transition_amp}
    \bra{f}T(t_0,t)\ket{i} = \left(\frac{-i}{\hbar}\right)^2 \sum_n \int_{t_0}^{t} dt'\: \bra{f}e^{i\:H_0t'/\hbar}H_{I}(t')e^{-i\:H_0t'/\hbar}\ket{n}\\ \int_{t_0}^{t'} dt''\: \bra{n}e^{i\:H_0t''/\hbar}H_{I}(t'')e^{-i\:H_0t''/\hbar}\ket{i}
\end{multline}
\end{center}
where the interaction Hamiltonian H\textsubscript{I\textsubscript{1}} 
gets contribution both from weak and electromagnetic interactions. 
We express this as
\begin{equation}
	H_I = H_{I_1} + H_{I_2}
\end{equation}
where $H_{I_1}$ and $H_{I_2}$ are respectively the weak and electromagnetic contributions. The weak interaction part, corresponding to the Fermi transition
being considered here, 
 is proportional to \cite{wong},
\begin{equation}\label{eq:weak_Hi}
    H_{I_1} \propto G_V \tau_-
\end{equation}
where $\tau_-$ is the isospin lowering operator and $G_V$ is the weak interaction coupling constant \cite{wong}. It is given by
\[G_V/(\hbar c)^3 = 1.149\times 10^{-11} {\rm MeV}^{-2} \]
The weak interaction Hamiltonian also contains terms which lead to creation
of a neutrino and a positron. These are not explicitly shown in Eq. \ref{eq:weak_Hi}.

The electromagnetic perturbation can be expressed as, 
\begin{equation}
	H_{I_2} =
\sum_i{ Z_ie\over c m_i} \vec A(\vec r_i,t)\cdot \vec p_i + \sum_i {e g_i\over 2m_ic}\vec S_i\cdot \vec B(\vec r_i,t)
    \label{eq:Hint}
\end{equation}
where $Z_i$, $m_i$ and $\vec p_i$ are respectively the charge, mass and momentum vector of the particle $i$. Here we have also
included a magnetic moment coupling with $g_i$ and $\vec S_i$ representing
the $g$ factors and the spin operators respectively. 
In Eq. \ref{eq:Hint}, $\vec A$ is the electromagnetic field, given by,
\begin{equation}
    \vec A(\vec r,t) = {1\over \sqrt{V}} \sum_{\vec k}\sum_\beta c\sqrt{\hbar\over 2\omega} \left[a_{\vec k,\beta}(t) \vec\epsilon_\beta e^{i\vec k\cdot\vec r} +  a^\dagger_{\vec k,\beta}(t) \vec\epsilon^{\,*}_\beta e^{-i\vec k\cdot\vec r}\right]
\end{equation}
and $\vec B=\vec \nabla\times \vec A$.

As explained earlier, the proton converts into a neutron through a 
Fermi transition. Next, the neutron gets absorbed by the nucleus through
an electromagnetic interaction. We assume that this is an electric 
transition where the charged particles in the nucleus ${}^A$X emits a photon
while the neutron gets absorbed. A complete calculation is complicated since
it would require us to use the full multiparticle wave function of
the nucleus $X$. Here we assume that ${}^A$X is a point particle of charge $Ze$, which reduces the problem into an effective one body problem. 

\subsection{Resonant Contribution}
In this subsection we make a model for the case in which the intermediate state gets 
contribution from a resonance. We are interested in relatively low energies. At such 
energies, the neutron absorption cross section shows a $1/v$ behaviour where $v$ is
the velocity of the incident neutron \cite{krane}. This is expected since the 
transition matrix element is expected to behave as a constant and we obtain a $1/v$ from the
flux factor. Besides that we see very sharp resonances for many nuclei even at relatively low energies or
order 1-10000 eV \cite{TERADA2015118,Youinou,chary2014}. Using date available from BNL National Nuclear Data 
Center, one finds that ${}^{58}$Ni also displays a resonance at roughly 7 keV with a width of roughly 7 eV. 
As we shall see this can contribute substantially to the reaction being considered. 
Here we are interested in modelling the phase shifts in the vicinity of a 
resonance. 

It is convenient to consider the standard square well potential with $V=-V_0$ for $r<r_0$
and $V=0$ for $r>r_0$. We consider the intermediate state wave function for $l=0$. We can express it as
$\chi_n = N_n u_n(r)/r$, where $N_n$ is the normalization factor. The wave function $u_n(r)$ for the square well for wave number $k_n$ is
given by,
\begin{eqnarray}
	u_n &=&  {A\over k_n}\sin(k_{n1} r)\  \ \ \ \ \ \ \ \ \ \ \ \ \ \ \ \ \ \ \ \ \ \ \   {\rm for}\ r<r_{0}\nonumber\\ 
	u_n &=&  {B\over k_n}\sin(k_{n} r) + {C\over k_n}\cos(k_n r)\ \ \ \ \ {\rm for}\ r\ge r_{0} 
\end{eqnarray}
where $k_{n1}^2 = k_n^2 + 2mV_0/\hbar^2$ and    
\begin{eqnarray}
	B&=& A \left[\sin (k_{n1}r_0) \sin (k_nr_0) +{k_{n1}\over k_n} \cos (k_{n1}r_0) \cos (k_nr_0) \right]\nonumber\\
	C&=& A \left[\sin (k_{n1}r_0) \cos (k_nr_0) -{k_{n1}\over k_n} \cos (k_{n1}r_0) \sin (k_nr_0) \right]
\label{eq:coeffpot}
\end{eqnarray}
The coefficient $A$ is fixed by setting $B^2+C^2 = 1$. We use plane wave normalization and 
at large $r$, $u_n(r) \rightarrow \sin(k_nr + \delta(k_n))/k_n$ where $\delta(k_n)$ is the
phase shift. As $k_n\rightarrow 0$, $A\rightarrow k_n$ and hence the nuclear matrix
element becomes independent of $k_n$, as expected. Furthermore $B$ and $C$ are proportional
to a constant and $k_n$ respectively. 

 Here we directly 
directly model $\delta(k_n)$ assuming presence of a resonance. Let us assume that
the phase shift can be modelled exactly by Eq. \ref{eq:reso}, ignoring higher order terms. 
We obtain,
\begin{eqnarray}
	\sin{\delta} &=& {-k_n\over \sqrt{k_n^2 + (a_1-a_2 k_n^2)^2}}\nonumber\\
	\cos{\delta} &=& {a_1-a_2k_n^2\over \sqrt{k_n^2 + (a_1-a_2 k_n^2)^2}}
\label{eq:resodelta}
\end{eqnarray}
In terms of the phase shift, $u(r) = \sin(k_nr + \delta)/k_n$ for $r\ge r_0$. 
This leads to the model, $C=-A/b $ and $B=A(a_1-a_2k_n^2)/(bk_n)$ close to resonance with
\begin{equation}
	A = {bk_n\over \sqrt{k_n^2 + (a_1-a_2k_n^2)^2}} 
\end{equation} 
where $b$ is a constant which is chosen so that this leads to nuclear matrix element identical to the
one obtained by the shell model potential (Eq. \ref{eq:potentialX})
in the limit $k_n\rightarrow 0$. 
These have exactly the same behaviour in the limit $k_n\rightarrow 0$ 
as the corresponding coefficients in the 
potential model, given in Eq. \ref{eq:coeffpot}. Hence in this case also the nuclear 
matrix element goes as constant as $k_n\rightarrow 0$.  
Our model for the phase shift displays the correct behaviour at very low energies
and in the vicinity of a resonance. We will use this to obtain the resonance contributions
for the calculation at second order in perturbation theory.

\subsection{Weak interaction perturbation}
The initial proton state $I_3=1/2$ while the intermediate neutron state
has $I_3 = -1/2$, where $I_3$ is the third component of the isospin operator.  
Let the integral over $t''$ in eq \ref{eq:transition_amp},
which corresponds to the amplitude for the first vertex in 
Fig. \ref{fig:process},  
be represented by $I_{2}$. We obtain,
\begin{equation}\label{eq:weak integral}
    I_2=\int_{t_0}^{t'} dt''\: \bra{n}e^{i\:H_0t''/\hbar}H_{I_1}(t'')e^{-i\:H_0t''/\hbar}\ket{i}
\end{equation}
The positron and the neutrino wave functions are given by $\exp(i\vec k_e\cdot \vec r)/\sqrt{V}$ and $\exp(i\vec k_\nu\cdot \vec r)/\sqrt{V}$ respectively. 
The operator $\tau_-$ leads to the transition between the $T_3=1/2$ proton
state to $T_3=-1/2$ neutron state. 
Hence we obtain
\begin{equation}\label{eq:weak integral1}
	I_2= {G_V\over V}\int_{t_0}^{t'} dt''\: e^{i(E_e +E_\nu + E_n-E_i)t''/\hbar} \int d^3r d^3R \psi^*_n(\vec R)\chi^*_n(\vec r) e^{-i\vec K\cdot \vec r_2
	}\psi_i(\vec R)\chi_i(\vec r)	
\end{equation}
where, $E_i$, $E_n$, $E_e$ and $E_\nu$ are the energies of initial state, 
the intermediate nuclear state, the emitted positron and the 
emitted neutrino respectively. Note that the rest mass difference in proton and neutron is included in the difference $E_n-E_i$. 
Furthermore, 
\begin{equation}
	\vec K= \vec k_e+\vec k_\nu
	\label{eq:Kvector}
\end{equation}
is the sum of the wave vectors of positron and neutrino. Integrating over time, we obtain
\begin{equation}\label{eq:weak integral2}
	I_2= -{i\hbar G_V\over V}\: {e^{i(E_e +E_\nu + E_n-E_i)t'/\hbar}\over E_n+E_e+E_\nu-E_i} \tilde I_2 
\end{equation}
where, 
\begin{equation}\label{eq:weak integralp}
	\tilde I_2=  \int d^3r d^3R \psi^*_n(\vec R)\chi^*_n(\vec r) e^{-i\vec K\cdot \vec r_2
	}\psi_i(\vec R)\chi_i(\vec r)	
\end{equation}
and we have dropped the term depending on $t_0$ since it will vanish in the limit $t_0\rightarrow-\infty$. Here $\psi_n(\vec R)\chi_n(\vec r)$ represents the
intermediate state wave function. 
We assume that the center of mass dependence 
of mass dependence is a plane wave, i.e.,
\begin{equation}
	\psi_n(\vec R) = {1\over \sqrt{V}} e^{i\vec K_n\cdot \vec R}
\end{equation}
where $\vec K_n$ is the wave vector corresponding to the center of mass. 
Setting $\vec r_2= \vec R + m_1\vec r/M$ we obtain
\begin{equation}
	\tilde I_2 = {(2\pi)^3\over V} \delta^3(\vec K_n+\vec K) I'_2
	\label{eq:I2tilde}
\end{equation}
where we have set $\psi_i(\vec R)=1/\sqrt{V}$ and   
\begin{equation}
	I'_2= \int d^3r \chi^*_n(\vec r)
	e^{-i\vec K\cdot \vec rm_1/M} \chi_i(\vec r)
	\label{eq:I2p}
\end{equation}

\subsection{Transition Matrix Element}
Besides the weak interaction, the transition matrix element also get 
contribution from the electromagnetic interaction. We can express it as,
\begin{equation}
	\langle f|T|i\rangle = -{1\over \hbar^2}\sum_n \int_{t_0}^t dt'
	e^{i(E_f-E_n)t'/\hbar} \langle f| H_{I_2}(t')|n\rangle I_2
\end{equation}
Here we get contribution from only the first term on the right hand
side of Eq. \ref{eq:Hint} 
if we ignore the magnetic moment of the neutron. The first term gets
contribution from the nucleus ${}^A$X with charge $Ze$ and mass $Am_N$. 
The final state wave function can be written as,
\begin{equation}
\Psi_f(\vec r,\vec R) = \psi_f(\vec R) \chi_f(\vec r) 
\end{equation}
The center of mass dependence is taken to be a plane wave
\begin{equation}
	\psi_f(\vec R) = {1\over \sqrt{V}} e^{i\vec K_f\cdot \vec R} 
\end{equation}

Let the wave vector of the emitted photon be
\begin{equation}
	\vec k_\gamma = k_\gamma[\cos\theta_2\hat z' + \sin\theta_2(\cos\phi_2
	\hat x'	+ \sin\phi_2 \hat y')]
\end{equation}
The photon field operator in Eq. \ref{eq:Hint} has the exponential factor
$e^{-i\vec k_\gamma\cdot \vec r_1}$. 
 This can be expressed as
\begin{equation}
	e^{i\vec k_\gamma\cdot \vec r_1} = e^{i\vec k_\gamma\cdot \vec R} 
	e^{-i\vec k_\gamma\cdot \vec rm_2/M}  
\end{equation}
Hence we obtain,
\begin{eqnarray}
	\langle f|H_{I_2}(t')|n\rangle &=& -i{Zem_2\over \sqrt{V}M} 
	\sqrt{1\over 2\hbar \omega}(E_f-E_n) e^{i\omega t'}\int d^3r' \chi^*_f(\vec r\,')
	\vec \epsilon_\beta\cdot \vec r{\,'} \chi_n\nonumber\\
	&\times &{(2\pi)^3\over V}\delta^3(\vec K_f+\vec k_\gamma-\vec K_n)	
	\label{eq:matrixfm}
\end{eqnarray}
Here we have set the exponential factor $\exp(-i\vec k\,'_f\cdot \vec r\,'m_2/M)\approx 1$ since the integral over $r'$ gets contributions only from very small
values of $r'$.

As mentioned earlier, for an order of magnitude estimate we can choose a 
particular final state. 
Hence we may consider only one of the two final state photon polarization
vectors. We take this to be, 
\begin{equation}
\vec \epsilon_\beta = -\sin\phi_2\hat x' + \cos\phi_2\hat y'
\end{equation}
The final state wave function $\chi_f$ is given in Eq. \ref{eq:wavefnX} with $m=1$.
\begin{equation}
	\chi_{f}(\vec r) = {u_X(r)\over r  } Y_1^1
\end{equation}
In order to proceed further we need to specify the intermediate state 
wave function $\chi_m$. Since the final state wave function has $l=1$ we
expect contributions both from $l=0$ and $l=2$. However, at low energies,
$l=2$ is expected
to be much suppressed in comparison to $l=0$ \cite{krane}. Hence we keep the contribution
only from $l=0$ which leads to,  
\begin{equation}
	\chi_n = {R_n\over \sqrt{V}} = {1\over \sqrt{V}} {u_n\over r}
	\label{eq:chim}
\end{equation}
where at large distances, $R_n(r) \rightarrow j_0(k_nr)$, up to a  
phase shift.  
We can project different harmonic components of the lepton
final state by
expanding the exponential factor in $I'_2$ (Eq. \ref{eq:I2p}), 
\begin{equation}\label{eq:partial wave}
e^{i\vec k\cdot \vec r} = 4\pi \sum_{l=0}^{\infty}
        \sum_{m=-l}^{l} i^l j_k(kr)\ Y_l^{m*}(\theta_k,\phi_k)
        Y_l^m(\theta_r,\phi_r)
\end{equation}
Only the $l=0$ component of the leptonic final state contributes. 

Evaluating the angular part of the integral in 
 Eq. \ref{eq:matrixfm}, we obtain,  
\begin{eqnarray}
	\langle f|H_{I_2}(t')|n\rangle &=& {Zem_2\over \sqrt{V}M} 
	{4\pi\over 3}\sqrt{3\over 16\pi E_\gamma} e^{-i\phi_2}
	(E_f-E_n) e^{i\omega t'}\int dr' r^{\prime\, 2} u_X \chi_n\nonumber\\ 
	&\times &{(2\pi)^3\over V}\delta^3(\vec K_f+\vec k_f-\vec K_n)	
	\label{eq:matrixfm1}
\end{eqnarray}
We therefore obtain, 
\begin{eqnarray}
	\langle f|T|i\rangle &=& i{G_V\over \hbar V}{Zem_2\over \sqrt{V}M}
	\sqrt{\pi
	\over 3E_\gamma}e^{-i\phi_2} \sum_n\int_{t_0}^t dt' 
	e^{i(E_f+E_\gamma+E_\nu+E_e-E_i)t'/\hbar}I_f{E_f-E_n\over
	E_\nu+E_e+E_n-E_i}I'_2\nonumber\\
	&\times & {(2\pi)^6\over V^2}\delta^3(\vec K_f+\vec k_\gamma -\vec K_n)
	\delta^3(\vec K_n+\vec K)
\end{eqnarray}
where 
\begin{equation}
	I_f = \int dr' r^{\prime 2}u_X \chi_n
	\end{equation}

The center of mass dependence is trivial. We need to take modulus square of the
transition amplitude 
and integrate over the final nuclear momenta after inserting the density
of states factor $V d^3K_f/(2\pi)^3$. This simply leads to a factor of unity
while imposing the momentum conservation,
\begin{equation}
\vec K_f + \vec k_\gamma + \vec K = 0
\end{equation}
Here we focus on the relative variable. 
The reaction rate is given by
\begin{equation}
	{dP\over dt} = {V^3\over \Delta T}\int {d^3k_\gamma d^3k_e d^3k_\nu 
	\over (2\pi)^9}|\langle f|T|i\rangle|^2 
\end{equation}
The integrand depends only on the magnitude, $K$, of the vector $\vec K$ (Eq. \ref{eq:Kvector}) and not on the directions of the positron or neutrino momentum
vectors. Hence for performing the integration over these momenta, we can
rotate our coordinate axis and choose the new $z$-axis to point along the
positron momentum vector. 
Let $\theta$ be the angle between the neutrino and positron momentum vectors. 
Hence we have, $K^2 = k^2_\nu+k^2_e + 2k_\nu k_e\cos
\theta$. 
This leads to, 
\begin{equation}
	{dP\over dt} = {V^3\over (2\pi)^9 \Delta T}
	\int dk_\gamma k^2_\gamma d\Omega_\gamma dk_e k^2_e d\Omega_e
	dk_\nu k^2_\nu 2\pi d\cos\theta 
\,	|\langle f|T|i\rangle|^2 
\end{equation}
The photon and positron angular integrals simply give a factor of $(4\pi)^2$. 
We convert the integral over $\cos\theta$ into an integral over $K$ and obtain,
\begin{equation}
	{dP\over dt} = {4V^3\over (2\pi)^6 \Delta T}
	\int dk_\gamma k^2_\gamma dk_e k^2_e
	dk_\nu k_\nu^2 \,  {KdK\over k_\nu k_e} 
\,	|\langle f|T|i\rangle|^2 
\end{equation}

A straightforward calculation gives,
\begin{equation}
	{dP\over dt} = {c\alpha G_V^2\over 6\pi^3 (\hbar c)^3}\left({Zm_2\over M}\right)^2
	\int dk_e dk_\nu dK k_e k_\nu K E_\gamma
	\Bigg|\sum_n I_f {E_f-E_n\over
        E_\nu+E_e+E_n-E_i} I'_2\Bigg|^2
\end{equation}
along with the energy conservation condition, 
\begin{equation}
E_f + E_\gamma + E_\nu + E_e - E_i = 0
\end{equation}
The integral over $K$ ranges from $|k_e-k_\nu|$ to $|k_e+k_\nu|$ where
$k_e$ and $k_\nu$ are the magnitudes of the positron and neutrino wave 
vectors respectively. Here we confine ourselves to the case where dominant contribution
is obtained
from relatively low intermediate state energies. The maximum value we will consider is about 10 keV. 
Hence the energy numerator, 
$E_f - E_n \approx -B$, where $B$ is the  
binding energy of the outer most neutron in the final state nucleus. This will be taken to be of order MeV. 
The energy denominator can be approximated as,
\begin{equation}
	E_\nu + E_e +E_n - E_i \approx E_\nu + E_e + (m_n-m_p)c^2 
\end{equation}
Since both the energy denominator and numerator are approximately independent
of the 
intermediate state energy, we can take them out of the sum over $n$.
This leads to,
\begin{eqnarray}
	{dP\over dt} &=& {c\alpha G_V^2\over 6\pi^3 (\hbar c)^3}\left({Zm_2\over M}\right)^2
	\int dk_e dk_\nu k_e k_\nu \int_{|k_e-k_\nu|}^{|k_e+k_\nu|} dK K 
	E_\gamma\nonumber\\
	&\times &\left[ {B\over E_\nu+E_e+(m_n-m_p)c^2}   \right]^2
	\Bigg|\sum_n I_f 
	I'_2\Bigg|^2
\end{eqnarray}

The sum over $n$ in the above equation can be expressed as
\begin{equation}
	\sum_n I_f I'_2 = {4\pi V\over (2\pi)^3} \int dk_n k_n^2 I_fI'_2
\end{equation}
Using Eq. 
\ref{eq:chim}, we obtain 
\begin{equation}
	\sum_n I_f I'_2 = {2N_i\over \pi} \int dk_n k_n^2
	\left(\int dr' r^{\prime 2} u_XR_n(r')\right)
	\left(\int dr r^2 R_n(r) j_0(Kr)\tilde\chi_i(r) \right)
\end{equation}
where we have explicitly substituted the initial state wave function and 
\begin{equation}
\tilde \chi_i = 
e^{-(r - r_0)^2/\Delta^2}
\end{equation}
For small $k_n$, the integral,
\begin{equation}
	I_f =  \int dr' r^{\prime 2} u_XR_n \approx -7.6\times 10^{-11}
	\label{eq:intib}
\end{equation}
in atomic units using the potential model given Eq. \ref{eq:potentialX}.  
This is applicable for the nickel nucleus within the framework of shell model potential. For 
other nuclei we expect a result with same order of magnitude assuming that a state exists with 
binding energy of same order as in the case of nickel.  
We take the numerical factor out of the $k_n$ integral and obtain,
\begin{equation}
	\sum_n I_f I'_2 = {2N_i\over \pi}(-7.6\times 10^{-11}) J 
\end{equation}
where $J$ is the integral  
\begin{equation}
	J = \int {k_n\over K}dk_n {a_1\over \sqrt{k_n^2 + (a_1-a_2k_n^2)^2}} 
	\int dr \sin(k_nr+\delta) \sin(Kr)\chi(r) 
	\label{eq:intJ}
\end{equation}
and $\delta$ is the phase shift. The factor $a_1/\sqrt{k_n^2 + (a_1-a_2k_n^2)^2}$ arises due to the intermediate
state wave function in the nuclear matrix element. This is normalized to unity in the limit $k_n\rightarrow 0$. 
The amplitude for the full process in governed by this integral
over $k_n$. 

\section{Results}
\begin{figure}[h]
\begin{center}
\includegraphics[ clip,scale=0.8]{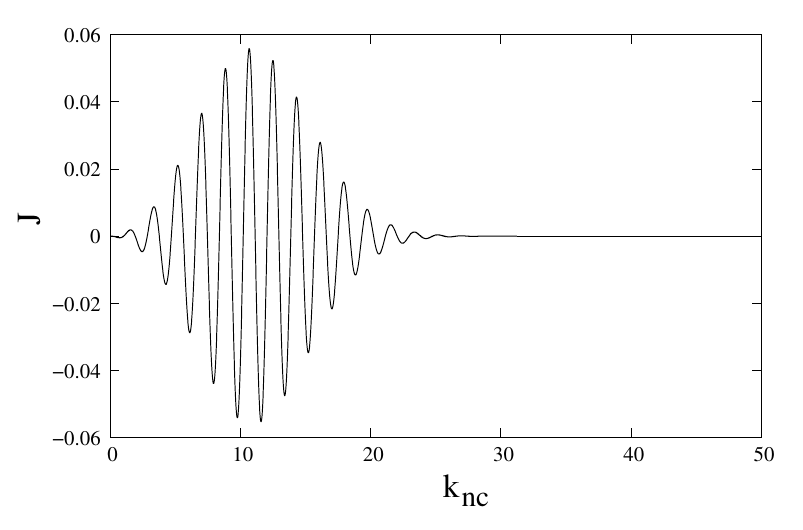} 
	\caption{\label{fig:cutoff1} The integral $J$ (Eq. \ref{eq:intJ}) as a function of $k_{nc}$, 
	the upper limit on the integration variable $k_n$ for the model $\delta=ak_n$, where $\delta$ is
	the phase shift and $a=0.1$ is a constant.} 
\end{center}
\end{figure}

We first study the integral $J$ for different models of the phase shift $\delta$.
We consider a simple model for the phase shift $\delta \approx a k_n$, where $a$ is a relatively small 
constant. Such a behaviour is expected for small values of $k_n$. In this case we find that the integral $J$
is very small. In Fig. \ref{fig:cutoff1} we show $J$ (Eq. \ref{eq:intJ}) as a function 
of the upper limit $k_{nc}$ on the integration variable $k_n$. Here we have set $a=0.1$ and $K=10$.  
 We find that although the integral 
becomes substantial for some values of $k_{nc}$, 
this integral approaches zero as $k_{nc}$ becomes large, leading to an extremely small 
rate.

\begin{figure}[H]
\begin{center}
\includegraphics[ clip,scale=0.8]{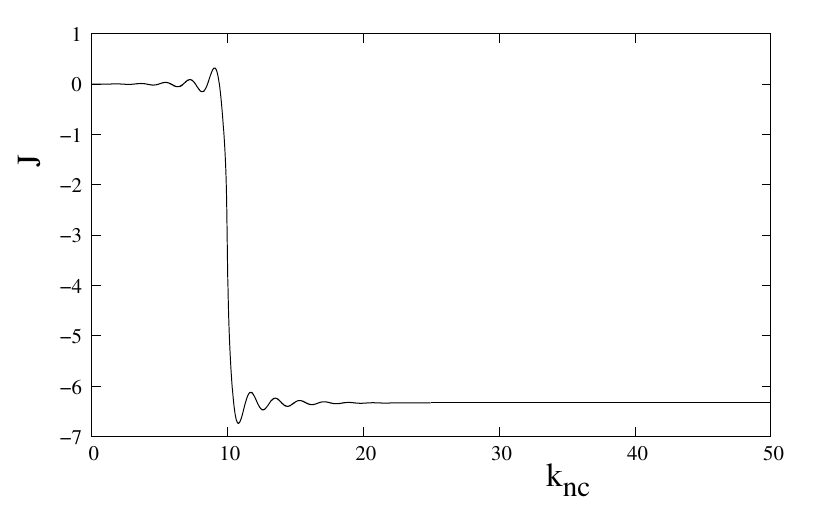} 
	\caption{\label{fig:cutoff2} The integral $J$ (Eq. \ref{eq:intJ}) as a function of $k_{nc}$, 
	the upper limit on the integration variable $k_n$ assuming a resonance at low energy.} 
\end{center}
\end{figure}

\begin{figure}[H]
\begin{center}
\includegraphics[ clip,scale=0.8]{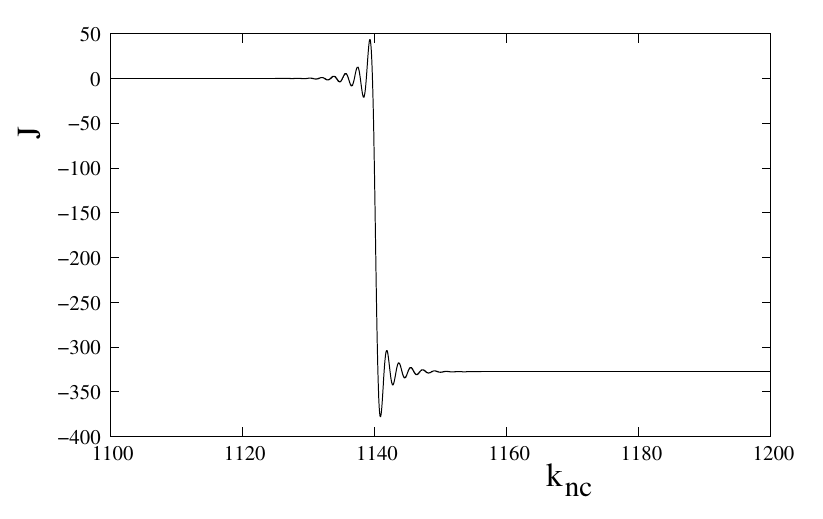} 
	\caption{\label{fig:cutoff3} The integral $J$ (Eq. \ref{eq:intJ}) as a function of $k_{nc}$, 
	the upper limit on the integration variable $k_n$ assuming a resonance at medium energy.} 
\end{center}
\end{figure}

We next consider the case when we are close to a resonance and use the model given in Eq. \ref{eq:resodelta}. 
We first consider a resonance at small $k_n$. We find that several nuclei show resonances in neutron absorption at
energies as small as 1 eV, see for example \cite{Youinou}. We set $a_1= 800$ and $a_2 = 8$ in Eq. \ref{eq:resodelta},
which leads to a resonance, with a relatively narrow width, at $k=10$.
In terms of energy, the resonance lies at $E_R=0.74$ eV with a width of 0.02 eV. 
The integral $J$ as a function of
the upper cutoff on $k_n$ is shown in Fig. \ref{fig:cutoff2}. Here we have set $K=10.4$. We see that the integral
saturates at a relatively large value. We find that $J$ takes relatively 
large values for $K$ less than roughly 11 and slowly
decays for large $K$.
In Fig. \ref{fig:cutoff3} we display results for another case in which the resonance is present at a higher energy.
In this case we choose $a_1 = 2.17\times 10^6$ and $a_2= 1.67$. Hence the resonance is located at $k= 1140.2$ or
$E_R= 9.6$ keV and has a
relatively narrow width of 10 eV. Here we have set $K = 1140.6$. We again see a substantial contribution.

In order to compute the rate we ignore the neutrino mass and use the relativistic formula 
for the positron
energy. For the parameters $a_1=800$, $a_2=8$, the rate is found to be $4\times 10^{-26}$
per second. This is found assuming that the nuclear matrix element is given by Eq.  
	\ref{eq:intib} which corresponds to nickel nucleus and a binding energy of 9 MeV. 
In general, for other nuclei this serves as
an order of magnitude estimate, assuming that the final state has sufficiently large binding energy 
so that the process is exothermic. We find that the rate is sufficiently large to be observable. The energy
spectrum of photons shows a sharp peak at approximately 7 MeV. This is essentially determined by energy conservation.

We next consider a case with resonance at higher energy, analogous to the result shown in Fig. \ref{fig:cutoff3}. 
We make small adjustment to the parameters so that the resonance corresponds to the one seen in 
nickel with resonance energy
of 7 keV and width of 7 eV. This can be obtained by setting 
$a_1 = 1.9\times 10^6$ and $a_2= 2.0$. In this case we find a much larger rate of about 
$10^{-16}$ per second. The peak in the photon spectrum is found to be around 3.5 MeV.
Hence we find a relatively large rate for the case of conversion of ${}^{58}$Ni to ${}^{59}$Ni which is 
 observable in laboratory. 
For example, in an electrochemical experiment we may be able to achieve of
the order of, or larger than, $10^{16}$ molecules composed of p-Ni. This would 
lead to a rate of one event per second which can be observed by detecting
the emitted photon at the energy peaked about 3.5 MeV.
 Similar rates are expected for other nuclei. 
A detailed comparison with experimental results is postponed to future research. 

We point out that a substantial contribution is obtained only in cases where the width of the 
resonance approximately matches the energy scale of oscillations seen in the absence of resonance
as a function of the intermediate state energy eigenvalue, as seen in Fig. \ref{fig:cutoff1}. 
If these two energy scales are very different the rate is expected to be very small. 
Furthermore, the resonance mechanism may be applicable to a wide range of
processes and not just the weak interaction process considered here. The
rates in these cases are expected to be much higher since they will not
be subjected to the weak interaction suppression.

\section{Conclusion}
In this paper we have considered a nuclear process in which an incident proton reacts with a target heavy nucleus X
to form an isotope of X with emission of a positron, neutrino and a photon. The process gets dominant contribution
at second order in perturbation theory. At the first vertex the proton converts into a virtual neutron through a weak
process. At the second interaction the neutron gets absorbed by the nucleus X with emission of a photon.
We find that the rate for such a process is negligible unless it is assisted by the presence of a nuclear resonance.
In the absence of resonance, we find significant contributions to amplitude from many intermediate neutron energy
eigenstates. However this contributions oscillate about zero and the sum is found to be negligible. In order to get a 
substantial contribution, 
the resonance width has to match the energy scale of the oscillations. Assuming the presence of such a resonance, 
we find that the rate is substantial and observable. We argue that this condition can be met in many nuclei observed in 
nature. We have directly applied our model to ${}^{58}$Ni which shows a resonance at approximately 7 keV with a
relatively narrow width of 7 eV. The rate in this case is found to be approximately $10^{-16}$ per second which 
is experimentally observable. 
Hence the proposed mechanism can be tested in laboratory. 
The resonance mechanism is also applicable to many other processes 
and not just confined to the weak process being considered here.
A detailed application to different processes and other nuclei as well
as comparison with experimental data is postponed to future research.

\section{Appendix}
In this equation we consider the Dirac equation for the isospin doublet, 
proton-neutron system and perform its non-relativistic reduction. The wave
function in isospin space is given by Eq. \ref{eq:isodoublet}. 
The Dirac equation can be written as
\begin{equation}
	i\hbar{\partial\over \partial t}
 \begin{bmatrix}
\psi_p \\ 
\psi_n
\end{bmatrix}
	= \left[c\vec \alpha\cdot \vec p 
	+\beta m_Nc^2 - {1\over 2} \beta\Delta mc^2\tau_3 + V\right]
 \begin{bmatrix}
\psi_p \\ 
\psi_n
\end{bmatrix}
\end{equation}
Here the first two terms on the right hand side are isospin symmetric,
and proportional to the
identity matrix in the isospin space, while the last two terms break this
symmetry. We point out that the potential $V$ is also a $2\times 2$ matrix
in isospin space. We denote its components by $V_{11}$, $V_{12}$, 
$V_{21} = V^*_{12}$ and $V_{22}$. 
We denote the upper and lower components in the Dirac space by
the symbols $\phi$ and $\zeta$ respectively. Hence for the proton, the 
equation becomes
\begin{equation}
	i\hbar{\partial\over \partial t}
 \begin{bmatrix}
\phi_p \\ 
\zeta_p
\end{bmatrix}
	= \left[c\vec \alpha\cdot \vec p 
	+\beta m_Nc^2 - {1\over 2}\beta\Delta mc^2 + V_{11}\right] 
 \begin{bmatrix}
\phi_p \\ 
\zeta_p
\end{bmatrix}
+ V_{12}
\begin{bmatrix}
\phi_n \\
\zeta_n
\end{bmatrix}
\end{equation}
The corresponding equation for the neutron is,
\begin{equation}
	i\hbar{\partial\over \partial t}
 \begin{bmatrix}
\phi_n \\ 
\zeta_n
\end{bmatrix}
	= \left[c\vec \alpha\cdot \vec p 
	+\beta m_Nc^2 + {1\over 2}\beta\Delta mc^2 + V_{22}\right] 
 \begin{bmatrix}
\phi_n \\ 
\zeta_n
\end{bmatrix}
+ V_{21}
\begin{bmatrix}
\phi_p \\
\zeta_p
\end{bmatrix}
\end{equation}
We next make the following transformation for the proton and
neutron wave functions
\cite{itzykson2012quantum}, 
\begin{equation}
	\phi_{p,n} = e^{-im_Nc^2t/\hbar}\Phi_{p,n}\ \ \ \ ,\ \ \  
	\zeta_{p,n} = e^{-im_Nc^2t/\hbar}Z_{p,n}
\end{equation}
The equations for the upper and lower components, $\phi_p$ and $\zeta_p$ can
be written as
\begin{eqnarray}
	i\hbar{\partial\over \partial t}\Phi_p  
	&=& c\vec \sigma\cdot \vec p\, Z_p 
	- {1\over 2}\Delta mc^2\Phi_p + V_{11} \Phi_p + V_{12}\Phi_n
	\label{eq:Phip}\\
	i\hbar{\partial\over \partial t}Z_p  
	&=& c\vec \sigma\cdot \vec p\, \Phi_p -2m_Nc^2 Z_p 
	+ {1\over 2}\Delta mc^2 Z_p + V_{11} Z_p + V_{12}Z_n
	\label{eq:Zp}
\end{eqnarray}
We now make the standard leading order approximation. Using Eq. \ref{eq:Zp}
we obtain, 
\begin{equation}
	Z_p \approx { \sigma\cdot \vec p \over 2c m_N}\Phi_p
\end{equation}
Substituting this in Eq. \ref{eq:Phip}, we obtain the
following leading order equation
for $\Phi_p$ in the non-relativistic limit, 
\begin{equation}
	i\hbar{\partial\over \partial t}\Phi_p  
	= {p^2\over 2m_N}\Phi_p 
	- {1\over 2}\Delta mc^2\Phi_p + V_{11} \Phi_p + V_{12}\Phi_n
	\label{eq:Phip1}
\end{equation}
Following similar steps, we obtain the corresponding equation for $\Phi_n$,
\begin{equation}
	i\hbar{\partial\over \partial t}\Phi_n  
	= {p^2\over 2m_N}\Phi_n 
	+ {1\over 2}\Delta mc^2\Phi_n + V_{22} \Phi_n + V_{21}\Phi_p
	\label{eq:Phin1}
\end{equation}

\bibliographystyle{unsrt}
\bibliography{Weak_interaction}
\end{document}